\documentclass[aps,pra,showpacs,notitlepage,twocolumn,longbibliography]{revtex4-1}
\usepackage{hyperref}
\usepackage{amsmath}
\usepackage{graphicx}

\newcommand{\be}{\begin{equation}}
\newcommand{\ee}{\end{equation}}

\newcommand{\bs}{\boldsymbol}

\begin{document}



\title{Artificial flat band systems: from lattice models to experiments}

\author{Daniel Leykam, Alexei Andreanov, and Sergej Flach}

\affiliation{Center for Theoretical Physics of Complex Systems, Institute for Basic Science (IBS), Daejeon 34126, Republic of Korea}

\begin{abstract}
Certain lattice wave systems in translationally invariant settings have one or more spectral bands that are strictly \emph{flat} or independent of momentum in the tight binding approximation, arising from either internal symmetries or fine-tuned coupling. These flat bands display remarkable strongly-interacting phases of matter. Originally considered as a theoretical convenience useful for obtaining exact analytical solutions of ferromagnetism, flat bands have now been observed in a variety of settings, ranging from electronic systems to ultracold atomic gases and photonic devices. Here we review the design and implementation of flat bands and chart future directions of this exciting field.
\end{abstract}

\maketitle
\section{Introduction}

Certain tight binding Hamiltonians have the peculiar property that one of more of their spectral bands are dispersionless, with a single particle energy spectrum $E(\bs{k})$ independent of momentum $\bs{k}$, forming \emph{flat bands}~\cite{derzhko2015strongly,parameswaran2013fractional,bergholtz2013topological}. The quenched kinetic energy in a flat band suppresses wave transport; the wave group velocity $\nabla_{\bs{k}} E$ vanishes. This results in a strong sensitivity to perturbations: any perturbation, no matter how weak, can set a new dominant energy scale for the flat band states and qualitatively change their transport properties. The generality of periodic media governed by Schr\"odinger-like equations $i \partial_t \psi = \hat{H} \psi$ allows flat bands to be explored in diverse settings from Hubbard models to Bose-Einstein condensates and photonics.

Perfectly flat bands are not stable against generic perturbations, which typically induce nonzero dispersion. For this reason, some authors broaden the definition of flat bands to include partially-flat bands that have vanishing dispersion only along particular directions or in the vicinity of special Brillouin zone points~\cite{cracknell1973van, deng2003the,nguyen2018}. In this review our focus is on bands that are perfectly flat, which require either fine-tuning of system parameters, or protection by a lattice symmetry. This limit of a perfectly flat band has provided useful settings for the study (and exact solution) of effects ranging from ferromagnetism to Anderson localization and superconductivity. 

It is quite remarkable that the original theoretical papers predicting the existence of flat bands are now over thirty years old (e.g. from 1986 for the dice lattice by Sutherland~\cite{sutherland1986localization}), but in many settings fabrication technologies have only recently caught up and achieved the precision required to realize artificial flat band lattices. For example, the Lieb lattice was originally introduced in 1989~\cite{lieb1989two}, received renewed interest from 2010 with optical lattice proposals~\cite{shen2010single,apaja2010flat}, and in the last few years the first experiments with cold atoms, photons, and electrons have finally been demonstrated. Many long-standing theoretical predictions of flat band phenomena may finally be tested in experiment. Now is therefore the ideal time to summarize and link the flat band literature from different fields, re-examine old predictions to see whether they can now be verified, and ask what effects should be targeted in future experiments. 

To these aims, this brief review is structured as follows: we begin in Section~\ref{sec:design} by recounting the history of how flat band models have been designed, starting from the study of isolated examples to a more systematic classification of their parameter space. We then discuss issues such as the behaviour of flat bands under perturbations, with emphasis on disorder, interactions, and topological phases. In the subsequent sections, we summarize the progress in realizing and probing artificial flat bands in different settings: electronic systems such as superconducting wire networks (Section~\ref{sec:electrons}), cold atoms in optical lattices (Section~\ref{sec:atoms}), and photonic systems including waveguide arrays and exciton-polariton condensates (Section~\ref{sec:photonics}). In the concluding Section~\ref{sec:conclusion} we will highlight the most recent theoretical advances which are promising to explore in future experiments, and ongoing topics of theoretical research. 

The focus of this review is on artificial flat band lattices created by structured potentials rather than the flat bands occurring in the atomic scale limit of magnetic materials. For a review of models of flat band magnetism in frustrated crystals, we recommend Refs.~\cite{ramirez1994strongly,chalker2011geometrically,derzhko2015strongly}.

\section{Designing and perturbing flat bands}
\label{sec:design}

The study of flat band models originates with Sutherland's observation~\cite{sutherland1986localization} of a flat band and its ``strictly localized states'' in the dice lattice (which relates to previous research on quasiperiodic lattices and Penrose tilings). Then came Lieb's seminal 1989 paper on the Hubbard model, where he proved that certain bipartite lattices with chiral flat bands exhibit ferrimagnetism at half filling, resulting in a macroscopic magnetization~\cite{lieb1989two}.

\begin{figure}
    \includegraphics[width=\columnwidth]{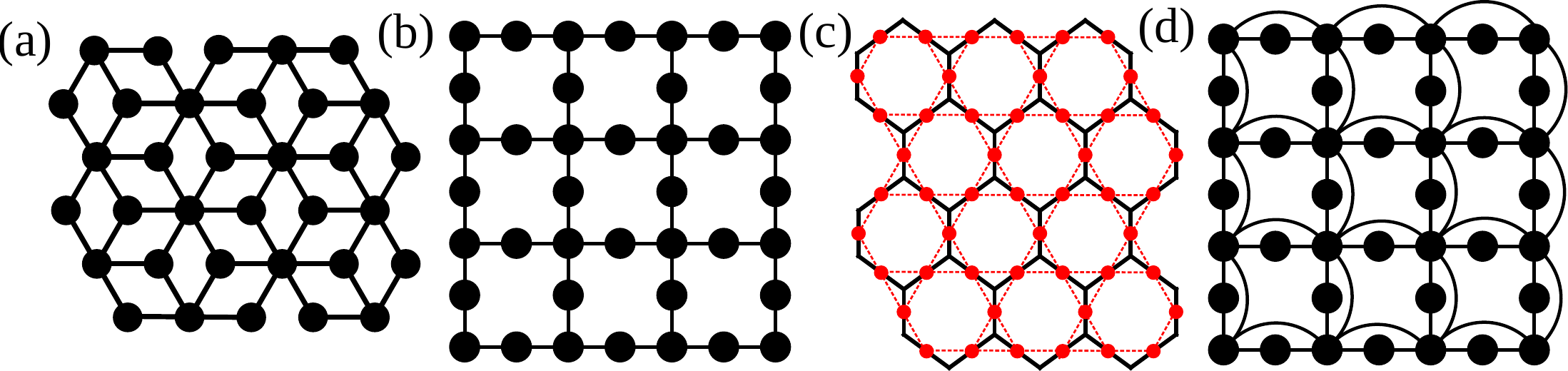}
    \caption{The first flat band lattices depicted as sites (filled circles) coupled via bonds (lines). 
        (a) Dice lattice \cite{sutherland1986localization}.
        (b) Lieb's lattice~\cite{lieb1989two}. 
        (c) Kagome lattice (red) as the line graph of the honeycomb lattice~\cite{mielke1991ferromagnetic,mielke1991ferromagnetism}. 
        (d) Tasaki's decorated square lattice~\cite{tasaki1992ferromagnetism,tasaki2008hubbard}.}
    \label{figure_1}
\end{figure}

Two examples of such a bipartite lattice are Sutherland's dice lattice shown in Fig.~\ref{figure_1}(a) and the edge-centered square lattice in Fig.~\ref{figure_1}(b), now known as the Lieb lattice. In both these examples, the sublattices can be divided into ``hub'' and ``rim'' sites. A bipartite symmetry emerges because the hub sites are only directly coupled to rim sites, and vice versa. 

Shortly afterwards Mielke and Tasaki independently generalized this idea to examples of flat band ferromagnetism, occurring when the lowest single particle energy band is flat~\cite{mielke1991ferromagnetic,mielke1991ferromagnetism,tasaki1992ferromagnetism,tasaki2008hubbard}. 
Mielke's construction of flat band lattices was based on line graphs, which are formed by promoting the links of an ordinary lattice into sites. For example, Fig.~\ref{figure_1}(c) illustrates how the Kagome lattice emerges as the line graph of the honeycomb lattice. On the other hand, the flat bands in Tasaki's ``decorated'' lattices emerge due to competition between fine-tuned nearest- and next-nearest neighbour hopping, see Fig.~\ref{figure_1}(d). 

\begin{figure}[t]

\includegraphics[width=\columnwidth]{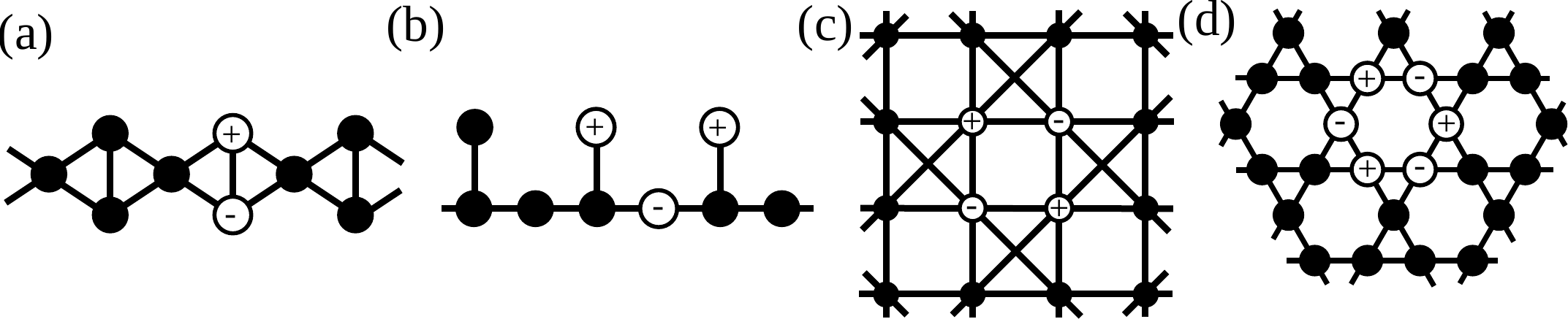}

\caption{Examples of compact localized eigenstates (CLS) of flat band lattices. White circles indicate the sites excited by the CLS along with the signs of the eigenstate amplitudes required to eliminate hopping to neighbouring unexcited sites. The flat band class $U$ denotes the number of unit cells occupied by the CLS. (a) Diamond ladder, $U=1$. (b) Stub lattice, $U=2$. (c) Checkerboard lattice, $U=3$. (d) Kagome lattice, $U=4$.}

\label{fig:CLS}

\end{figure}

Tasaki subsequently proved the stability of the magnetic order under weakly lifted degeneracy of the flat band, an important step in establishing that these flat band models were not pathological and could help understand real materials~\cite{tasaki1994stability}.
In all cases the researchers quickly realised that the macroscopic degeneracy of the flat band allows construction of compactly localized Wannier-like eigenstates from linear combinations of extended Bloch states, as shown in Fig.~\ref{fig:CLS}.

Subsequent theoretical studies of flat band ferromagnetism have focused on weakening the assumptions behind these rigorous results, e.g. by considering different filling factors and other classes of flat band lattices (see review~\cite{derzhko2015strongly}). The macroscopically degenerate ground states of ``frustrated'' lattices, where the lowest band is flat, host intriguing spin liquid phenomena~\cite{anderson1973resonating,balents2010spin}, an active area of research~\cite{kitaev2003fault,kitaev2006anyons,jackeli2009mott,savary2017quantum}, and exotic excitations such as magnetic monopoles~\cite{castelnovo2008magnetic,morris2009dirac,bramwell2009measurement,fennell2009magnetic}. Such frustrated lattices can also be fabricated artificially~\cite{wang2006artificial}, providing a useful tool to probe and understand their properties~\cite{nisoli2013colloquium}.

\subsection{Different flat band classes and generators}

In 1993 Shima and Aoki introduced a class of superlattices exhibiting symmetry-protected chiral flat bands~\cite{shima1993electronic}, which have received renewed interest from 2012~\cite{lan2012coexistence,zhong2017transport}. A later paper by Aoki et al. was the first (to our knowledge) to introduce the name ``compact localized state'' (CLS) for flat band eigenmodes constructed as superpositions of the degenerate Bloch waves~\cite{aoki1996hofstadter} (note that the term ``strictly localized state'' introduced by Sutherland in 1986~\cite{sutherland1986localization} and used in the context of quasiperiodic Penrose lattices~\cite{kohmoto1986electronic,arai1988strictly,nori1990angular} did not spread in the community).

Aoki et al. observed that the above flat band models are divided into two classes based on their response to an applied magnetic field~\cite{aoki1996hofstadter}. In Mielke's and Tasaki's lattices a magnetic field destroys the fine-tuned interference responsible for the flat band and breaks it into a Hofstadter butterfly spectrum. In contrast, sublattice symmetry-protected chiral flat bands such as in the dice and Lieb lattice do not depend on the precise values of the coupling strengths, but only on the ``local topology'' of the sublattice connectivity: the rim sites forming a ``majority sublattice'' hosting the flat band are only connected via an intermediate  ``minority'' sublattice, the hub sites. Consequently, chiral flat bands remain macroscopically degenerate under an applied magnetic field, which only shifts the energies of a few of the flat band states~\cite{sutherland1986localization,nita2013spectral}.

In 2008, the idea of classifying flat bands through the properties of their compact localized eigenmodes emerged. Bergman et al.~\cite{bergman2008band} showed that the set of CLS in two dimensions can turn linearly dependent precisely when the flat band touches a non-flat (dispersive) band at one or more points in the Brillouin zone. This linear dependence implies that the basis formed by the CLS is incomplete and does not span all states belonging to the flat band. The missing states are line states, which are delocalized (Bloch wave-like) along one direction, and compactly-localized in the perpendicular direction. In the case of chiral flat bands, an additional extended state also resides on the minority sublattice~\cite{ramachandran2017chiral}. 

Since 2014, compact localized states were used to construct ``generators'' of flat band networks, which goes beyond the original construction approaches of Mielke and Tasaki. Rather than identifying flat bands in particular lattice models, one can assume a flat band with a given CLS exists and obtain a family of lattice Hamiltonians compatible with this constraint. The size of the CLS $U$ - the number of unit cells the CLS occupies - serves as an important parameter of the generators.

Whenever the CLS can be localized to a single unit cell of the lattice ($U=1$), the flat band states can be completely decoupled from the rest of the lattice by a local change of basis, i.e. the flat band is protected by a local symmetry associated with this transformation~\cite{flach2014detangling}, which also presents the most general flat band generator for these CLS types in any lattice dimension. 

On the other hand, when the CLS occupy multiple unit cells, the location of flat bands with respect to other dispersive bands becomes constrained; the flat band arises due to fine-tuning, rather than a local symmetry. The parameter space of these flat bands ($U=2$ cells) assuming short-ranged hopping was determined systematically in 2017~\cite{maimaiti2017compact}. Higher $U$ flat bands can also be generated from their compact localized states by solving an inverse eigenvalue problem, suggesting a systematic way to obtain flat bands with a desired class $U$. 

Compact localized state-based construction methods have so far been limited to one-dimensional lattices; a complete generalization to higher dimensions remains an open problem. In 2017 Ramachandran et al.~\cite{ramachandran2017chiral} introduced the most general flat band generator for bipartite lattices with chiral flat bands, where the first example (to our knowledge) of a chiral flat band in a three-dimensional network was obtained. Other novel approaches for obtaining flat band lattices involve origami rules~\cite{dias2015origami}, the repetition of oligomers~\cite{weimann2016transport}, local symmetries~\cite{roentgen2018compact}, and self-similar (fractal) constructions~\cite{nandy2015engineering,pal2017flat}. 

\subsection{Disorder and interactions}
 
Different classes of flat bands respond qualitatively differently to perturbations. One can understand their behaviour by considering the projection of the perturbation to the flat band subspace. In the simplest $U=1$ case the projected perturbation acts locally, but is resonantly enhanced for eigenmodes close to the flat band energy. Roughly speaking, this effective resonant enhancement occurs because the perturbation sets the only relevant energy scale for the flat band states. When $U>1$ the projected interaction is no longer local, but decays exponentially with distance if the flat band is gapped~\cite{huber2010bose}, and with a power law if a symmetry forces the flat band to touch a dispersive band edge~\cite{chalker2010anderson}. 

One important class of perturbations is disorder, which induces Anderson localization in regular (non flat band) lattices. In a flat band the renormalization of the effective disorder potential transforms disorder distributions with finite variance into heavy-tailed distributions with diverging 
variance, modifying the scaling of the localization length in 1D~\cite{leykam2013flat,leykam2017localization,ge2017anomalous}. Mobility edges can also be obtained if the disorder potential has correlations that preserve the shape of the CLS~\cite{bodyfelt2014flatbands}.

In two dimensions, the long range decay of the projected interaction induces multifractality of the flat band eigenstates in the weak disorder limit~\cite{chalker2010anderson}. An ``inverse'' Anderson transition has been demonstrated numerically by Goda et al. in three-dimensional disordered flat bands: all eigenstates are localized for weak disorder, and delocalize at a critical disorder strength, before localizing again at at a second transition at stronger disorder~\cite{goda2006inverse,nishino2007flat}. Such localization-delocalization-localization behaviour as a function of the disorder strength also occurs in the level spacing statistics of certain two-dimensional flat bands~\cite{shukla2017criticality}. In the past year flat bands under non-quenched (evolving) disorder~\cite{radosavljevic2017light}, disorder-induced topological phase transitions~\cite{chen2017disorder}, and the temporal dynamics of disordered flat band states~\cite{gneiting2018} started to attract attention too.

A second active area of theoretical study is the dynamics of interacting quantum particles in flat bands, where interactions can induce spontaneous symmetry breaking and a variety of strongly-correlated quantum phases~\cite{zhao2012quantum,takayoshi2013phase,tsai2015interaction,metcalf2016matter,luo2016gauge,bercx2017spontaneous,chen2017disorder}. The vanishing wave group velocity in flat bands simplifies numerical studies by guaranteeing that even weak interactions are capable of producing rich phase diagrams. On the other hand, exact ground states and analytical solutions can be obtained at certain filling factors~\cite{huber2010bose,takayoshi2013phase}. Exact solutions where interactions induce flat dispersion in an otherwise dispersive band have also been demonstrated~\cite{gulacsi2014interaction}.

Interestingly, the CuO$_2$ planes in cuprate high temperature superconductors have a Lieb lattice structure and it is conjectured that flat band effects may be responsible for their high critical temperature~\cite{kopnin2011high,iglovikov2014superconducting,peotta2015superfluidity,julku2016geometric,kobayashi2016superconductivity,tovmasyan2016effective,liang2017band}. In ordinary dispersive bands the superfluid weight is inversely proportional to the effective mass. While this vanishes in a flat band due to the diverging effective mass, nevertheless interactions can induce ballistic transport due to nonzero overlaps between the Wannier functions~\cite{tovmasyan2016effective}. This leads to additional contributions to the superfluid weight, that are not universal to all flat energy dispersions but sensitive to the geometrical and topological properties of the band~\cite{iglovikov2014superconducting,peotta2015superfluidity,julku2016geometric,liang2017band}. This geometrical contribution to the superfluid weight is proportional to the interaction strength, rather than the hopping strength as is the case for regular dispersive bands.

In the mean field approximation the interactions in flat bands can be described by discrete nonlinear Schr\"odinger equations. Nonlinearity breaks the superposition principle, nevertheless compact localized flat band modes can persist provided each site excited by the mode experiences the same nonlinear frequency shift. The nonlinear compact localized modes or solitons can be stable if they avoid resonance with other dispersive bands, e.g. if the flat band is gapped~\cite{gligoric2016nonlinear,zegadlo2017single}. In two-dimensional flat band lattices the soliton threshold power (nonzero in regular lattices) can vanish because solitons can bifurcate immediately from the compact localized states existing in the linear limit~\cite{vicencio2013discrete,belicev2017localized}. For this to occur, the weak nonlinearity must shift the CLS into a band gap; this mechanism does not work in lattices such as the 2D Lieb, where the flat band is embedded witin dispersive bands~\cite{lazarides2017squid}. It is also possible to use nonlinearity to generate compact localized states when no flat bands exist in the linear limit~\cite{johansson2015compactification}.

\subsection{Topological flat bands}

Another interesting direction is to use flat bands to study novel strongly interacting topological phases of matter. This requires models that exhibit flat or near-flat bands while simultaneously having nontrivial topological invariants. In one dimension this is relatively simple to achieve and was explored by Creutz as early as 1999~\cite{creutz1999end}. Another direction is to introduce flat bands to existing topological models, such as by imposing a bipartite symmetry, allowing the zero energy topological edge startes to interact with the zero energy flat band~\cite{bercioux2017}. Generalizing these ideas to higher dimensions has turned out to be a much more challenging task.

Wide interest in two-dimensional topological flat band models was sparked by a trio of back to back papers published in Physical Review Letters in 2011~\cite{tang2011high,sun2011nearly,neupert2011fractional} predicting room temperature fractional quantum Hall states in suitably-designed topological flat band models. Their idea was to start with a topologically nontrivial lattice Hamiltonian, and then optimize hopping parameters and on-site energies to maximise the ratio of the band width to band gap. When this ratio is sufficiently small interactions can induce fractional quantum Hall-like eigenstates. Such models are frequently referred to as fractional Chern insulators in the literature~\cite{parameswaran2013fractional,bergholtz2013topological}.

Despite intense theoretical activity on topological flat bands~\cite{weeks2012flat,guo2012fractional,budich2013fractional,parameswaran2013fractional,bergholtz2013topological}, including generalization of the original models to higher Chern numbers~\cite{wang2011nearly,wang2012fractional,yang2012topological}, these lattice fractional quantum Hall states have not yet been observed in experiment due to a number of challenges, both fundamental and practical.

First, optimization of the band flatness ratio is not a trivial task. It is trivial to obtain a perfectly flat band from any lattice model by dividing the Hamiltonian by the energy of one of its bands, but this will typically induce unphysical long range hopping terms. Truncating the hopping to a finite range often spoils either the nontrivial topology or the band flatness. It is therefore challenging to maintain a near-flat band while keeping the hopping short-ranged.

In fact, subsequent papers have revealed topological obstructions to achieving simultaneously nontrivial topology, perfectly flat bands, and short-range hopping; at best one can have only two of the three~\cite{chen2014the,read2017compactly,lee2016band}. Furthermore, while lattice fractional quantum Hall effects have been numerically demonstrated in a variety of models, if one wants exact analogues of the continuum fractional quantum Hall states, other quantities characterizing the band such as the Berry curvature and quantum metric should also be flat, giving additional less intuitive design constraints~\cite{bergholtz2013topological}. Last year, Lee et al.~\cite{lee2017band} introduced an efficient way to simultaneously optimize all these constraints using an elliptic function parametrization of the band's Bloch functions.

Despite the lack of simple topological flat band models in two or more dimensions, flat band-inspired ideas can still be useful for understanding the behaviour of topological surface states. Kunst et al. have designed exactly solvable models of topological surface and corner states~\cite{kunst2017anatomy,kunst2017lattice}, based on stacking layers and fine-tuning the interlayer coupling. The advantage of this approach is that, given the exact analytical solution for surface states, it is then straightforward to calculate their response to perturbations or interactions. 

For further reading related to topological flat bands, we recommend the recent review articles Refs.~\cite{parameswaran2013fractional,bergholtz2013topological}.

\section{Electronic flat bands}
\label{sec:electrons}

Until 1998 flat bands were largely studied in the context of spin models. That year an influential paper by Vidal et al.~\cite{vidal1998aharonov} found that in certain periodic electronic networks, such as the dice lattice, a critical magnetic field strength induces a completely flat spectrum~\cite{vidal1998aharonov}, in contrast to the complex Hofstadter butterflies that had previously been studied~\cite{nori1990angular,aoki1996hofstadter}. They called this effect ``Aharnonov-Bohm (AB) caging'' because the completely flat spectrum occurred when each unit cell was threaded by a $\pi$ magnetic flux, inducing interferences analogous to the Aharonov-Bohm effect. They proposed that this effect could be most easily observed in superconducting wire networks, where the required magnetic flux quantum is significantly reduced compared to atomic-scale systems (mT fields versus 10$^3$T).

\begin{figure*}
    \includegraphics[width=1.5\columnwidth]{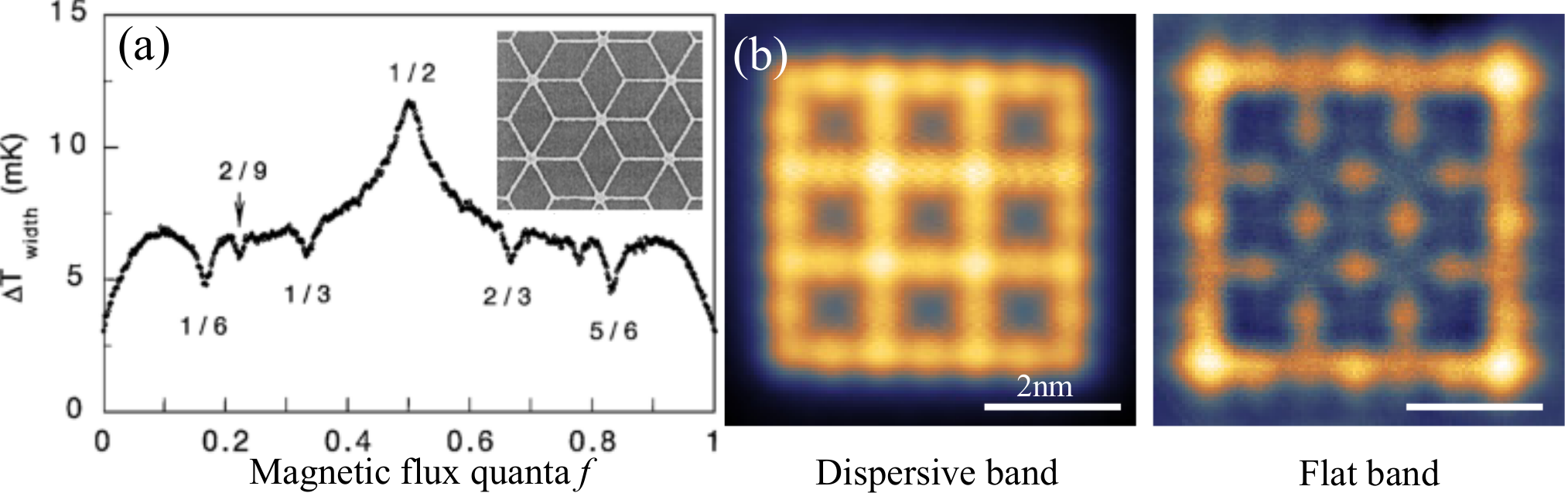}
    \caption{(a) Aharonov-Bohm cage in superconducting wire network. The superconducting transition width $\Delta T$ shows a broad peak in the flat band limit $f=1/2$, associated with the formation of decoupled compact localized states. Inset is an image of the dice lattice network~\cite{abilio1999magnetic}. (b) Selective imaging of dispersive and flat band states of an electronic Lieb lattice by changing the applied bias voltage and performing spatially-resolved conductance measurements~\cite{drost2017topological}.}
    \label{fig:electronic_FB}
\end{figure*}

\subsection{Superconducting networks}

In superconducting networks one cannot directly observe the single particle band structure and measurements are limited to transport properties such as the conductivity. Furthermore, one can only probe the ground state properties, requiring the lowest band to be flat and excluding lattices such as the Lieb. In 1999, following the proposal by Vidal et al.~\cite{vidal1998aharonov}, Abilio et al. reported the following indirect observations of flat bands in an AB cage~\cite{abilio1999magnetic}:

(i) Anomalies in the critical temperature $T_c \approx 1.24K$. $T_c$ measures the eigenvalue of the tight binding model's ground state. While anomalies typically occur at rational flux quanta, an especially strong anomaly reducing $T_c$ by a few percent occurs in the AB cage limit of a half a flux quantum~\cite{lin2002quantum}. 

(ii) A broadened superconducting transition width $\Delta T$, illustrated in Fig.~\ref{fig:electronic_FB}(a). Because the AB cage eigenmodes are compactly localized, the phases at different plaquettes are no longer correlated, resulting in a broadened transition width.

(iii) Vanishing critical current, which measures the band curvature at the ground state. The critical current typically peaks at rational values of the flux, except for the AB cage limit where it forms a minimum. The measured value of critical current was nonzero (17\% of zero field value), which Abilio et al. attributed to either edge states or interaction effects. 

Subsequent experiments in 2001 observed signatures of AB caging in the magnetoresistance oscillations of a low temperature (30mK) normal metal dice lattice~\cite{naud2001aharonov}. In both cases, the experiments used a lattice period of $\approx 1\mu$m to minimize the required magnetic field.

Theoretical studies have generalized AB caging to disordered and interacting networks~\cite{vidal2001disorder} and Josephson junction arrays~\cite{doucot2002pairing}. In the latter, transport in the AB cage limit can be induced by two-body interactions between Cooper pairs, resulting in a novel four-electron superconducting state~\cite{rizzi20064e,tesei2006frustration}. In two-dimensional systems, the suppression of single pair transport can be used to achieve superconducting qubits that are topologically protected from local noise by parity symmetry~\cite{douccot2003topological,douccot2005protected}. The 4e superconducting state was first observed in 2009~\cite{gladchenko2008superconducting}, and there are now efforts to scale these experiments to large two-dimensional arrays~\cite{bell2014protected}.

\subsection{Engineered atomic lattices}

Advances in fabrication of 2D materials have enabled engineering of nanoscale artificial lattices for electrons using techniques such as lithography and atomic manipulation~\cite{tadjine2016from,qiu2016designing}. In 2017, 2D Lieb lattices were embedded onto a substrate surface with a scanning tunnelling microscope (STM) using two different techniques~\cite{drost2017topological,slot2017experimental}. Drost et al. removed atoms from a chlorine monolayer, leaving the desired lattice structure~\cite{drost2017topological}. In contrast, Slot et al. added carbon monoxide molecules to a substrate, which similarly structured the repulsive potential felt by the surface electrons~\cite{slot2017experimental}, an idea independently proposed by Qiu et al.~\cite{qiu2016designing}.

In both cases the energy scale of the electron hopping is approximately $0.1$eV, and the STM allows the spatially-resolved measurement of the electron density through conductance spectroscopy. By changing the bias voltage one can also selectively measure either the flat or dispersive band Bloch waves, as shown in Fig.~\ref{fig:electronic_FB}(b). On the other hand, in the experiments to date the Fermi level is not tunable and does not coincide with the flat band. In the future, using heavier atoms it may also be possible to observe spin-orbit coupling effects in this platform. 

Evidence of a kagome lattice flat band has also been reported in a multilayer silicene structure imaged with an STM~\cite{li2017realization}. 

\section{Optical lattices with flat bands}
\label{sec:atoms}

In the cold atom community the Lieb lattice stands out as the prototypical model for exploring flat band phenomena in optical lattices. Not only does it host a wealth of novel effects when interactions are introduced, but it is also relatively simple to transfer atoms into the flat band. This distinguishes the Lieb lattice from earlier-studied lattices such as the dice~\cite{bercioux2009,bercioux2011} and kagome~\cite{santos2004,kagome_2}, the latter realized experimentally as early as 2012 by Jo et al.~\cite{jo2012ultracold}. For example, while the kagome lattice is easier to implement in experiment, only requiring interference between two triangular lattices, its flat band forms the highest excited state in the tight binding approximation. This makes it more challenging to probe the kagome lattice's flat band with cold atoms compared to the Lieb lattice.

\subsection{The beginning}

The first proposal for an optical Lieb lattice by Shen et al. in 2010~\cite{shen2010single} was based on six detuned standing wave laser beams. Changing the relative amplitude of the laser beams tunes the depths of the sublattices and can gap out one of the dispersive bands. In Ref.~\cite{shen2010single} the main interest was not in the flat band itself, but rather in the conical intersection formed by the dispersive bands~\cite{leykam2016conical}, which was shown to support wide-angle Klein tunneling of wavepackets through potential barriers.

\begin{figure}
    \includegraphics[width=\columnwidth]{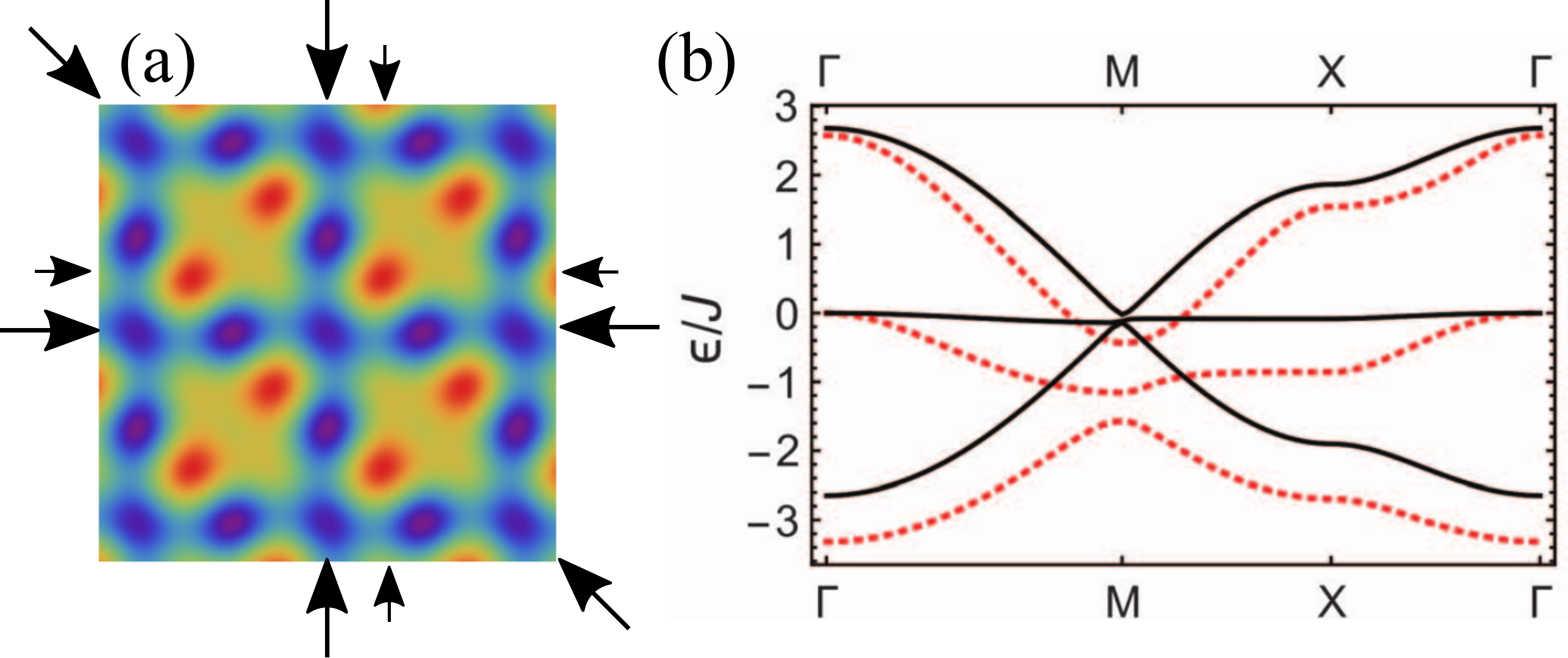}
    \caption{An optical Lieb lattice for cold atoms. (a) Lattice potential formed by five interfering standing waves (indicated by oppositely-aligned arrows). Lattice sites lie at the potential minima in blue. (b) Band structure for shallow (red dashed lines) and deep (solid black lines) lattices computed in Ref.~\cite{taie2015coherent}.}
    \label{fig:optical_lieb}
\end{figure}

One can see in Fig.~\ref{fig:optical_lieb} that in contrast to the ideal tight binding model of the Lieb lattice, in shallow or moderately-deep optical lattices the next-nearest neighbor hopping is typically nonzero, giving the ``flat'' band a finite width. Apaja et al. numerically computed the exact Bloch wave spectrum for this familiy of optical lattices, finding that the relative intensities of the interfering laser beams could be optimized to obtain an almost-flat band with width only 1.5\% of the total bandwidth~\cite{apaja2010flat}. Performing numerical simulations of wavepacket dynamics, they further showed that fermionic atoms could remain confined by the flat bands while repulsively interacting bosons typically tunnel to the dispersive bands and escape. Later studies found that these effects are also observable in a variety of quasi-1D flat band lattices~\cite{hyrkas2013many,metcalf2016matter}. 

\subsection{Experimental Lieb lattice} 

The first realization of an optical Lieb lattice for bosonic cold atoms was reported in 2015 by the group of Takahashi~\cite{taie2015coherent}. They found that only five standing waves are enough to produce a flat band if the lattice is made sufficiently deep.

In their experiment they used dynamic tuning of the optical lattice to transfer the condensate from the ground state into the flat band. First, the depth of the corner sites (minority sublattice) was reduced to transfer the condensate into the rim sites (majority sublattice). Secondly, an imbalance between the horizontal and vertical rim sites was used to imprint a $\pi$ phase difference between them. Quickly resetting the lattice to the ideal Lieb lattice configuration, the final state corresponds to a flat band eigenstate.

Consistent with the prediction by Apaja et al.~\cite{apaja2010flat}, interactions were observed to induce a decay of the condensate into the lower dispersive band. While gapping the flat band (by detuning the hub sites) can suppress this interband tunneling, the limiting factor to the measured flat band lifetime of $\sim0.5$ms was the condensate decay, which is insensitive to the gap size. 

The group of Takahashi has subsequently published two further studies, both in 2017. In the first, Ozawa et al.~\cite{ozawa2017interaction} used a weak external force to move the condensate through the Brillouin zone and map the band structure by measuring the local group velocity of the lowest band and the gaps to higher bands. Interactions were found to shift the energy of the flat band at the edge of the Brillouin zone, resulting in nonzero dispersion. These measurements were consistent with the tight binding limit of the Gross-Pitaevski equation. 

The second study by Taie et al. implemented for the first time fermionic cold atoms in the Lieb lattce and demonstrated a dark state-mediated adiabatic transport of particles between the horizontal and vertical rim sites, analogous to stimulated Raman adiabatic passage~\cite{taie2017spatial}. To date these are the only experimental studies of the Lieb lattice flat band with cold atoms.

Other optical lattice flat band lattice geometries are also starting to attract interest. For example, in 2017 An et al.~\cite{an2017flux} realized a flat band in a quasi-1D sawtooth lattice with an effective magnetic flux. 

\subsection{Future directions}

Experiments have so far been limited to single component (spinless) atoms. While novel topological Mott insulating phases have been predicted for spinless Rydberg atoms with long range interactions~\cite{dauphin2016quantum}, perhaps the most interesting future direction is to extend the present experiments to spinful atoms.

Using multi-component atoms it is possible to implement the synthetic gauge fields originally proposed by Goldman et al. in 2011~\cite{goldman2011topological}, based on laser assisted tunnelling between different hyperfine states. The resulting spin-orbit coupling is predicted to gap the flat band while preserving its flatness, inducing a topological insulating phase with helical edge states similar to that studied by Weeks and Franz~\cite{weeks2010topological}. In contrast to the fractional Chern insulator models of Refs.~\cite{tang2011high,sun2011nearly,neupert2011fractional}, here the flat band remains perfectly flat and topologically trivial; it is the dispersive bands that become topologically nontrivial.

A similar assisted tunneling scheme can also implement an Aharonov-Bohm cage on the dice lattice, which has the advantage that the flat band is the lowest band rather than an excited state~\cite{moller2012correlated}. These synthetic gauge fields can also potentially induce strongly interacting topological phases in one-dimensional flat bands~\cite{junemann2017exploring}.

A second opportunity offered by spinful atoms is to observe the long-predicted flat band ferromagnetism. This will require cooling of the atoms to even lower temperatures; while recent experiments in simple square and cubic lattices have started to observe evidence of antiferromagnetic ordering, the temperatures achieved so far remain above the critical temperature $T_c$ at which magnetically-ordered phases emerge~\cite{hart2015observation,drewes2017antiferromagnetic}. Promising however is a recent prediction that $T_c$ may be enhanced in flat bands~\cite{noda2015magnetism}.

In the above two dimensional lattices, an additional confining potential along the third dimension is required to trap the atoms. Noda et al.~\cite{noda2014flat,noda2015magnetism} considered the case where this confining potential is also periodic, resulting in a multilayer Lieb lattice. Using dynamical mean field theory, they showed that flat-band ferromagnetism occurs for odd layer numbers. The ferromagnetic ground state can advantageously be detected by measuring the magnetisation of the different layers, rather than different sublattices. Therefore the transition to an antiferromagnetically-ordered state is potentially easier to measure using multilayer optical lattices.

\section{Photonic flat bands}
\label{sec:photonics}

In photonics flat bands are closely related to the technologically-important concept of slow light~\cite{baba2008slow}, where the suppression of the wave group velocity offers enhanced nonlinear effects and is useful for pulse buffering. In applications, one always has the trade-off between reducing the group velocity, and maintaining a useful bandwidth of operation (delay-bandwidth product constraint), with ideal ``flat bands'' corresponding to a diverging delay with a vanishing bandwidth. 

An early proposal by Takeda et al. in 2004 for achieving flat band models was based on photonic crystal slabs of high-index dielectric rods, which exhibit band structures well-described by the tight binding approximation~\cite{takeda2004flat}. At the time, however, it was far easier to fabricate ``inverse'' structures based on air holes in high index substrates, which cannot be easily mapped to tight binding lattices, and this proposal was not successfully pursued. 

Instead, the design of slow light structures was larely focused on 1D waveguides or 2D lattices formed by defects in triangular photonic crystals optimized using numerical or intuitive approaches~\cite{baba2008slow,li2008systematic,xu2015design}. In 2017 Schulz et al. found that flat band model-inspired alternatives to the standard triangular structures, such as kagome lattices, can offer improved group velocity reduction~\cite{schulz2017photonic}. 

Interest in flat band structures was reinvigorated in 2010 with proposals for achieving tight binding bands in plasmonic waveguide networks~\cite{feigenbaum2010resonant,endo2010tight}, and with the experimental demonstration of geometric frustration in a kagome lattice coupled laser array~\cite{nixon2013observing}. Flat bands in kagome~\cite{nakata2012observation} and Lieb~\cite{kajiwara2016observation} structures were later realized for terahertz spoof plasmons, and there is now a significant body of literature implementing flat band models in dielectric waveguide arrays.

\begin{figure*}
    \includegraphics[width=1.5\columnwidth]{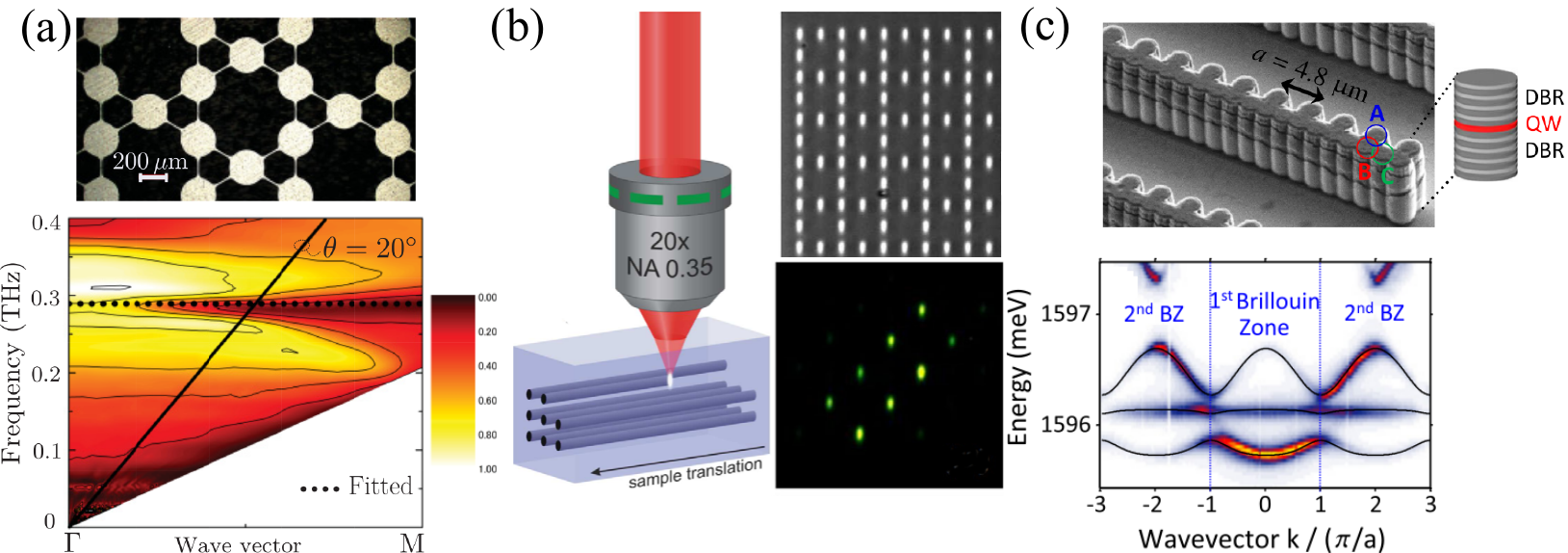}
    \caption{Examples of photonic flat bands: (a) Kagome lattice for terahertz spoof plasmons, displaying an omnidirectional minimum in the transmission at the flat band frequency (dashed line) ~\cite{nakata2012observation}. (b) Femtosecond laser-written Lieb lattice waveguide arrays and an observed compact localized flat band state~\cite{vicencio2015observation}. (c) Structured microcavity forming a 1D stub lattice and its photoluminescence spectrum revealing a middle flat band~\cite{baboux2016bosonic}.}
    \label{fig:photonic_FB}
\end{figure*}

\subsection{Femtosecond laser written waveguide arrays}

The femtosecond laser-writing technique has been used to fabricate dielectric waveguide arrays since 2004~\cite{szameit2010discrete}, but flat band phenomena have only become accessible in the last few years with the development of aberration-correction techniques that allow precise fabrication of two-dimensional arrays of sufficiently deep waveguides. The advantages of femtosecond laser writing over other photonic platforms is the ability to achieve long dimensionless propagation distances combined with near-arbitrary control over the inter-waveguide coupling, allowing exploration of effects such as time-periodic (Floquet) lattice modulations, and Bloch oscillations induced by weak potential gradients. 

The first demonstration of a two-dimensional Lieb lattice was reported in 2014~\cite{guzman2014experimental}, but the single waveguide input in this original experiment excited a superposition of all the bands, so the presence of the flat band could only be inferred indirectly. The following year Vicencio et al. and Mukherjee et al.~\cite{vicencio2015observation,mukherjee2015observation} used a phase modulated, multi-waveguide input beam to directly excite the flat band's CLS and observe their nondiffracting behaviour. 

Since then, several further experiments have been reported in various lattices~\cite{mukherjee2015observation1,weimann2016transport,diebel2016conical,maczewsky2017observation,mukherjee2017experimental,mukherjee2017observation,real2017all,cantillano2017observation}, studying quasi-1D flat bands~\cite{mukherjee2015observation1,weimann2016transport}, flat bands induced by periodic driving~\cite{maczewsky2017observation,mukherjee2017experimental,mukherjee2017observation}, forming optical logic gates out of CLS~\cite{real2017all}, and using superlattice structures to minimize the detrimental effect of next-nearest neighbour coupling on the band flatness~\cite{diebel2016conical,cantillano2017observation}.

\subsection{Optically-induced lattices}

Optically-induced lattices are created by applying a lattice-writing beam to a photorefractive medium, which results in modulation of the refractive index proportional to the lattice beam's intensity. Because of its very different characteristics, the optical induction platform is complimentary to the laser-writing technology, capable of accessing effects that are difficult to achieve in fused silica.

Whereas the femtosecond laser-writing technique is based on permanently damaging the host glass, lattices created by the optical induction technique can be written and erased at will, making significantly easier to obtain large ensemble averages (e.g. of disordered systems). Nonlinear wave dynamics are observable with low power continuous wave laser beams. However, arbitrary waveguide geometries are not possible, and the waveguides tend to be shallower and more anisotropic (more like a photonic crystal than a discrete lattice). The accessible dimensionless propagation distances are also significantly shorter, both due to the smaller crystal sizes (2cm compared to up to 10cm for fused silica), and the weaker effective coupling strength required to minimize detrimental next-nearest neighbour coupling.

The first observations of flat bands in photorefractive crystals were published in 2016~\cite{xia2016demonstration,zong2016observation}. The 2D Lieb lattice was generated using a superposition of mutually incoherent square lattices~\cite{xia2016demonstration}, similar to the methods used for optical lattices for cold atoms, while the Kagome lattice can be generated with a single induction beam~\cite{zong2016observation}. A different approach used to induce several quasi-1D flat band lattices originally popularized by Hyrk\"as et al.~\cite{hyrkas2013many} was based on incoherent superpositions of non-diffracting Bessel beams, with each Bessel beam inducing a single waveguide~\cite{travkin2017compact}.

\subsection{Polariton condensates}

Similar to the above wavewguide lattices, polariton condensates are described by a (2+1)D Schr\"odinger equation. Experiments with a microcavity exciton-polariton condensate in a frustrated kagome lattice potential were originally reported in 2012~\cite{masumoto2012exciton}, where photoluminescence was used to meausure the single particle band structure. At the time fabrication methods were limited to relatively shallow potentials, requiring a trade-off between keeping the lattice period small enough so that its band structure could still be resolved, and minimizing the next-nearest neighbour interactions which induce nonzero dispersion. Furthermore, as the flat band is not the ground state, condensation in the flat band could not be observed into these experiments.

With improved fabrication techniques such as etching of micropillar cavities it is now possible to implement ideal flat bands in their tight binding limit. In 2014 Jacqmin et al. made the first observation of polariton flat bands in the photoluminescence spectrum of the P orbital bands of a honeycomb array~\cite{jacqmin_FB}. Subsequently, Baboux et al used a 1D stub lattice combined with spatially structured pumping to achieve polariton condensation into a flat band~\cite{baboux2016bosonic}. Due to intrinsic disorder and suppression of transport in the flat band, they observed multiple condensates corresponding to different CLS instead of a single coherent condensate. These experiments only considered relatively weak pump powers, at which polariton-polariton interaction effects could be neglected. Going to higher powers, this platform is promising for the study of mean field nonlinear effects such as flat band solitons.

In contrast to waveguide arrays, polariton condensates can display strong spin-orbit coupling due to TE-TM splitting, which leads to a spin-dependent hopping strength. At the single particle level this results in CLS with nontrivial spin textures in symmetry-protected flat bands such as the 2D Lieb~\cite{whittaker2018exciton,klembt2017} and is predicted to induce nonzero dispersion in line graphs such as the kagome lattice~\cite{gulevich2016kagome}. In the interacting regime the spin-orbit coupling enables the emulation of spin chain models. Recent experiments have observed the formation of tunable ferromagnetic and antiferromagnetic ordering in coupled condensate chains~\cite{ohadi2017spin}, as well as frustration in 2D arrays of up to 45 spins~\cite{berloff2017realizing}. Further increases in the array size offer the exciting prospect of quantum simulation of systems intractable on classical computers.

\section{Outlook}
\label{sec:conclusion}

The last few years have seen significant experimental advances in our ability to engineer and probe flat bands for electrons, cold atoms, and photons. Specifically, we now have the ability to fine-tune lattices to create the desired flat band lattice geometry, and to probe the lattices with high resolution to spatially resolve the sublattice structure and compact localized states. Linear properties of the flat bands have now been observed in all these settings, and in platforms such as cold atoms and polaritons attention is now focused on observing novel quantum, nonlinear and interaction effects to pursue very recent theoretical predictions such as novel superconductivity. 

There are a few notable platforms where flat bands have been proposed, but not yet demonstrated in experiment: optomechanical arrays~\cite{wan2017controllable} and cavity QED systems at microwave and optical frequencies~\cite{qed_review}. Dissipation and interactions in these settings are predicted to induce transport and novel quantum correlations of the flat band states~\cite{biondi2015,schmidt2016frustrated,yang2016circuit,casteels2016probing,owen2017dissipation,rota2017on,biondi_arxiv}. This is an area ripe for experimental studies. Similar quantum multi-photon flat band states may also be explored in optical waveguide arrays~\cite{rojas-rojas2017quantum}.

External fields perturb flat bands and lead to strongly anharmonic Bloch oscillations of wave packets \cite{khomeriki2016landau}. Applying DC fields
in dimension $d=2$ turns a flat band into an infinite Wannier-Stark ladder set of one-dimensional flat bands with nontrivial topology \cite{kolovsky2018topological,long2017topological}. The details of the impact of the DC field direction, and the extension to dimension $d=3$ are still waiting to be explored. Perturbed compact localized states from a flat band act as Fano scatterers for dispersive waves and can be useful for spectroscopy~\cite{ramachandran2018fano}.

The driven-dissipative nature of systems such as polaritons and coupled-cavity arrays offers the interesting prospect of studying non-Hermitian effects in flat bands. There have been a few theoretical proposals on this topic. The first class considered effect of adding non-Hermiticity to existing Hermitian flat bands, e.g. by replacing each site in a flat band lattice by non-Hermitian dimers with balanced gain and loss~\cite{ge2015parity,chern2015pt,zhang2016dispersion,molina2015flat,park2018quantum}. In these examples, it was found that either the existing compact localized states become amplified or attenuated, or the compact localized states are destroyed by unflattening of the energy spectrum~\cite{qi2018defect}.

In 2017 two theoretical studies proposed flat bands induced by a non-Hermitian potential: one based on fine-tuning of gain and loss in a sawtooth chain~\cite{ramezani2017non}, and the other based on non-Hermitian coupling that preserves a bipartite sublattice symmetry~\cite{leykam2017flat}. In both cases, the non-Hermitian flat bands coincided with non-Hermitian degeneracies embedded in a continuum; the flat band was not isolated but is immersed in a dispersive band. This however does not appear to be a generic situation, as the combination of non-Hermitian on-site potentials and coupling terms allows for isolated non-Hermitian flat bands~\cite{ge2018non}.

So far only ideal non-Hermitian flat bands are considered in detail; but typically perturbations will break the balance between gain and loss and induce preferential amplification of certain modes, which will dominate in the resulting dynamics. This is a topic that needs to be understood further, as it will have an important impact on future experiments. 

A second outstanding issue raised by the case of non-Hermitian Hamiltonians is what is the most appropriate definition of a flat band. One has in principle three alternatives: both the real and imaginary parts of the band are flat, or just the real part~\cite{qi2018defect}, or just the imaginary part. Furthermore, in practical realizations one might only require the band flatness to hold locally rather than over the entire band~\cite{cracknell1973van, deng2003the,nguyen2018}.

In summary, advances in fabrication techniques make flat bands an increasingly exciting area of study. Models previously regarded as mere theoretical conveniences are now becoming accessible in experiments with electrons, atoms, and photons. This not only motivates the continued study of disordered, quantum, and strongly-interacting flat band systems, but sets the stage for harnessing flat band physics in future micro- and nano-scale devices.

\section*{Acknowledgement}

This work was supported by the Institute for Basic Science in Korea (IBS-R024-D1 and IBS-R024-Y1).

\bibliography{flatband,josephson,frustration}

\end{document}